\documentclass[3p,times]{elsarticle}

\usepackage{ecrc}

\volume{00}

\firstpage{1}

\journalname{Nuclear Physics A}

\runauth{Hard Probes PB Gossiaux}

\jid{nupha}

\usepackage{amssymb}

\usepackage[figuresright]{rotating}
\usepackage{here}
\usepackage{graphicx}

\begin{document}

\begin{frontmatter}



\dochead{}

\title{Recent results on heavy quark quenching in ultrarelativistic heavy ion collisions: the impact of 
coherent gluon radiation}


\author{Pol Bernard Gossiaux}
\address{SUBATECH, UMR 6457 (Ecole des Mines de Nantes, IN2P3-CNRS, Universit\'e de Nantes), 
4 rue Alfred Kastler, 44300 Nantes, France}

\begin{abstract}
We present a model for radiative energy loss of heavy quarks in quark gluon plasma which incorporates 
coherence effects. We then study its consequences on the radiation spectra as well as on the nuclear modification 
factor of open heavy mesons produced in ultrarelativistic heavy ion collisions.
\end{abstract}


\end{frontmatter}


\section{Introduction}
Ultimately, one expects to probe fundamental properties of the hot medium formed in 
ultrarelativistic heavy ions collisions (URHIC) thanks to the jet quenching observed in these URHIC. 
In this respect, the quenching of heavy quarks (HQ) enriches the analysis and permit to test some general 
concepts like mass-hierarchy~\cite{Dokshitzer:2001}. Understanding the nature of HQ energy loss is 
a rather broad and complex subject reviewed f.i. in~\cite{Peigne:09}. In \cite{Gossiaux:sqm09}, we have 
advocated that the gluon formation length characterizing the radiative energy loss is small enough at RHIC 
energies to evaluate this quantity through a generalization of the Gunion-Bertsch (GB) 
approach~\cite{Gunion:1981qs} for finite mass and have obtained good agreement with the data. 
As this ansatz is questionable at LHC, we have developed a model to study the 
impact of coherence on HQ quenching and present here our main results. We insist that all shocks 
at the origin of the radiation are treated in a running coupling approach~\cite{Gossiaux:08}. Therefore, the 
large value of $\alpha_s$ at small momentum transfer prevents, in our view, the direct application of most of the 
available frames designed in recent years for dealing with this topic.

\vspace{-0.3cm}
\section{Estimates for the formation times of 
radiated gluons}
\label{sect_estimates_lf}
As the regime pertinent to the gluon radiation depends on the comparison between scales pertaining to 
the medium and the gluon formation length $l_f$, we start by estimating this quantity. 
In the case of a gluon emitted through a single scattering, $l_f$ can be estimated from the 
virtuality of various Feynman diagrams contributing to the radiation \cite{Gunion:1981qs}.
\vspace{-0.3cm}
\begin{figure}[H]
\begin{center}
\includegraphics[height=3cm] {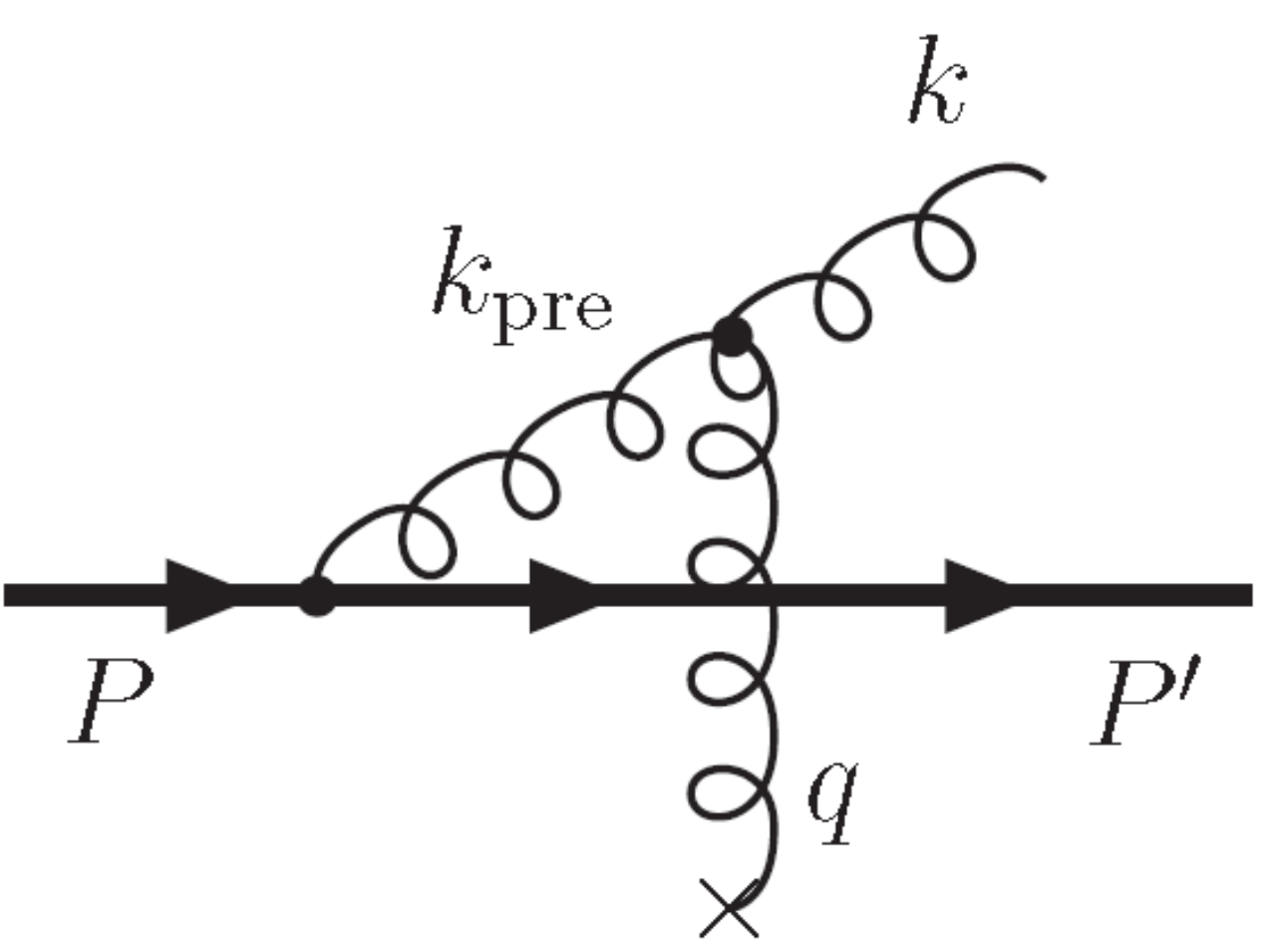}
\hspace{1cm} 
\includegraphics[height=3cm] {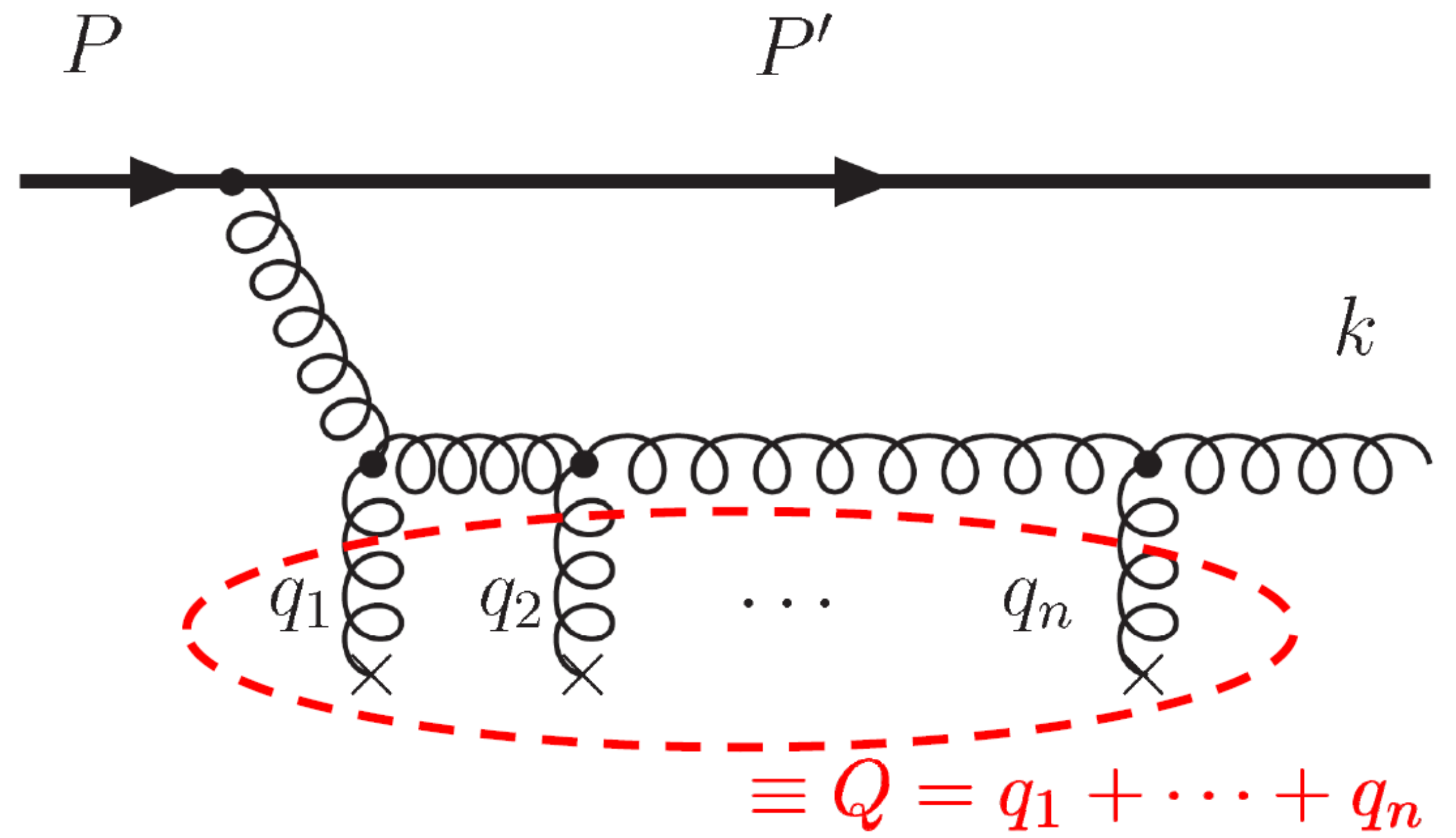} 
\end{center}
\vspace{-0.5cm}
\caption{3-gluons diagram of the $x+Q\rightarrow x'+Q'+g$ process (left) and typical diagram describing 
gluon radiation induced in a multiple scattering process (right).}
\label{fig_GB}
\end{figure}
For the case of the ``3-gluons'' diagram on fig. \ref{fig_GB} (left), specific of QCD and contributing 
to the radiation at mid-rapidity, one finds $t_f=\frac{2 k_{\rm pre}^0}{|k_{\rm pre}^2-m_g^2|}$,  
where $k_{\rm pre}=k-q$ is the 4-momentum of the pre-gluon and $m_g$ is the thermal gluon mass~\cite{Kampfer:00}
used in~\cite{Djordjevic:04b} for the evaluation of radiative energy loss along a finite opacity 
expansion. For $\gamma\gg 1$, one obtains
\begin{equation}
l_{f,{\rm sing}}^{3\,{\rm gl}}\approx t_f \approx 
\frac{2(1-x)\omega}{x^2 M^2+(1-x)m_g^2 +(\vec{q}_\perp -\vec{k}_\perp)^2}\,,
\end{equation}
where $M$ is the quark mass and $x$ the energy fraction of the gluon $x=\omega/E$.
When $M=m_g=q_\perp$, we recover the usual law $l_f\approx \frac{2\omega}{k_\perp^2}$
\cite{Baier:95}. As most of the shocks happen with $q_\perp\approx \mu$ (the Debye screening-mass associated 
to the medium) and lead to the radiation of a gluon with $k_\perp\approx \mu$ as well, one gets 
$\langle l_{f,{\rm sing}}\rangle \approx \frac{2(1-x)x E}{x^2 M^2+(1-x)m_g^2 + \mu^2}$,
with $m_g\sim \mu$ for practical purpose. At $x\approx x_{\rm cr}\triangleq \frac{\mu}{M}$, $l_f$ 
admits a maximum ${\rm max}(l_f)\approx \frac{E}{m_g M}$. Accordingly, coherence will affect mostly 
gluons radiated at $x\sim x_{\rm cr}$ and its impact will be reduced for HQ due to the 
hierarchy $l_f(M_u)>l_f(M_c)>l_f(M_b)$. 
Comparing $l_f$ with the mean free path $\lambda$, one obtains a first qualitative 
criterion that permits to estimate whether coherence effect should be included. For $T=0.25$ GeV and $\alpha_s=0.3$,
one finds that it is the case for $E_c>10~{\rm GeV}$ and for $E_b>25~{\rm GeV}$.

We now turn to the estimation of $l_f$ in a multi-scattering process happening inside a medium considered as 
infinite. Following the semi-classical argument of LP~\cite{Landau}, we proceed by considering the phase 
$\Phi(t)=\int_0^{+\infty} P(t')\cdot k(t') \frac{dt'}{E}$ associated to the radiation process, where $P$ and 
$k$ represent the instantaneous momentum of the parton and of the radiated gluon. Such a phase is also 
found in diagrammatic approach stemming from QFT~\cite{Baier:95}, up to recoil corrections for finite $x$. 
Assuming that the rescattering primarily affects the gluon due to its larger color charge, one finds 
\begin{equation}
\langle \Phi(l)\rangle\approx \frac{\omega}{2}\left[\left(\frac{M^2}{E^2}+\frac{m_g^2}{\omega^2}\right)l
+\frac{\hat{q}}{\omega^2} \frac{l^2}{2}\right]=\frac{l}{l_{f,{\rm sing}}}+\frac{l^2}{l_{f,{\rm mult}}^2}\,,
\label{eval_phi}
\end{equation}   
for $E\gg M$, where $\hat{q}$ is the gluon transport coefficient, that is the mean square momentum transfer per 
mean free
path\footnote{Including diagrams where the quark undergoes rescattering leads to an effective MFP 
$\tilde{\lambda}=2\lambda_g$ which should be taken to calculate the 
$\hat{q}$.} $\lambda_g$. The second term in the r.h.s reflects the 
contribution of the rescatterings to the increase of the phase, that is to the separation
of the gluon from its radiating parton. $l_f$ is then defined as the length for which 
$\Phi=\Phi_{\rm dec}\sim 1$. Accordingly, one gets $l_f\approx {\rm min}(l_{f,{\rm sing}},l_{f,{\rm mult}}
\triangleq 2\sqrt{\Phi_{\rm dec}\omega/\hat{q}})$. 
First, we note that the coherence effects mainly show up when the second
term dominates over the first, that is for $l_f\gtrsim L^{\star\star}\triangleq
\frac{l_{f,{\rm mult}}^2}{l_{f,{\rm sing}}} \approx\frac{m_g^2+ x^2 M^2}{\hat{q}}$. 
For finite $x$ and $M\gg m_g$, $L^{\star\star}\gg \lambda$,
meaning that coherence effects do not affect significantly the spectrum for $l_f\in[\lambda,
L^{\star\star}]$. Comparing  $l_{f,{\rm sing}}$ and $l_{f,{\rm mult}}$, one obtains the
following regimes\footnote{Further discussion can be found in \cite{Bluhm:12}.}, 
already identified in \cite{Peigne:09}: a) (low energy regime) for $E\lesssim 
E_{\rm no\; LPM}\approx M m_g^3/\hat{q}$, $l_{f,{\rm sing}}<l_{f,{\rm mult}}$
and coherence effects are thus negligible; b) (intermediate energy regime)
when $E\gtrsim E_{\rm no\; LPM}$, one has $l_{f,{\rm mult}}\lesssim l_{f,{\rm sing}}$ 
for $x\in [x_1,x_2]$ with $x_1\sim m_g^4/\hat{q}E$ and $x_2\sim\sqrt[4]{\hat{q}E^4/M^4}$
and coherence effects should be taken into account on this domain; c) (high energy regime) when 
$E\gtrsim E_{\rm LPM}\approx M^3/\hat{q}$, $x_2\sim 1$ and coherence effects dominate 
the full spectrum, except for a small domain at $\omega\lesssim \tilde{\lambda} 
\langle q_\perp^2\rangle\sim T$. In this large energy regime, the formation length becomes mass-independent and one
expects similar radiation for all flavors.

\section{Model for gluon radiation induced
by multiple collisions}
\label{section_model}
\begin{figure}[H]
\begin{center}
\includegraphics[height=3.cm] {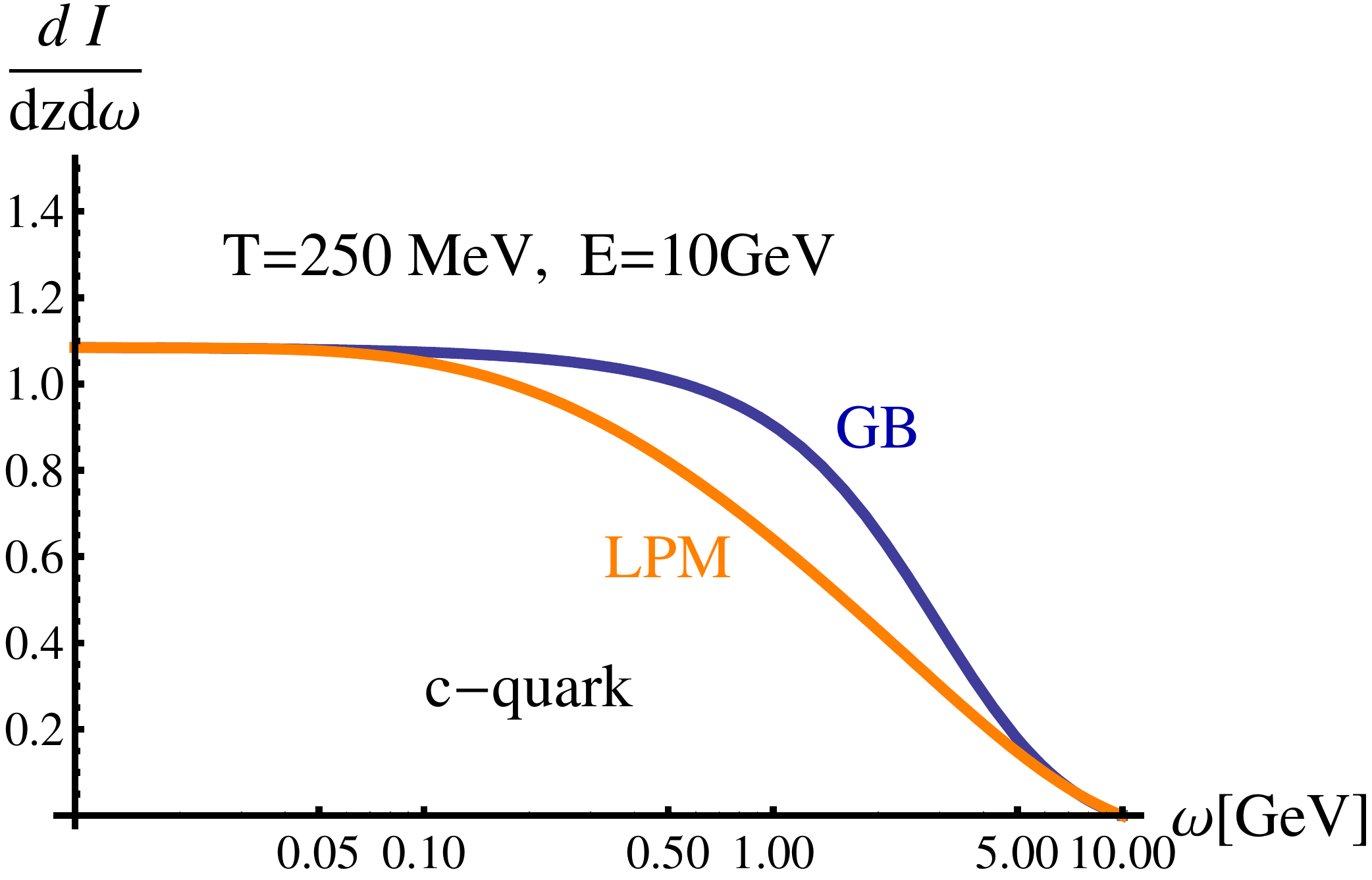}
\hspace{1cm} 
\includegraphics[height=3.cm] {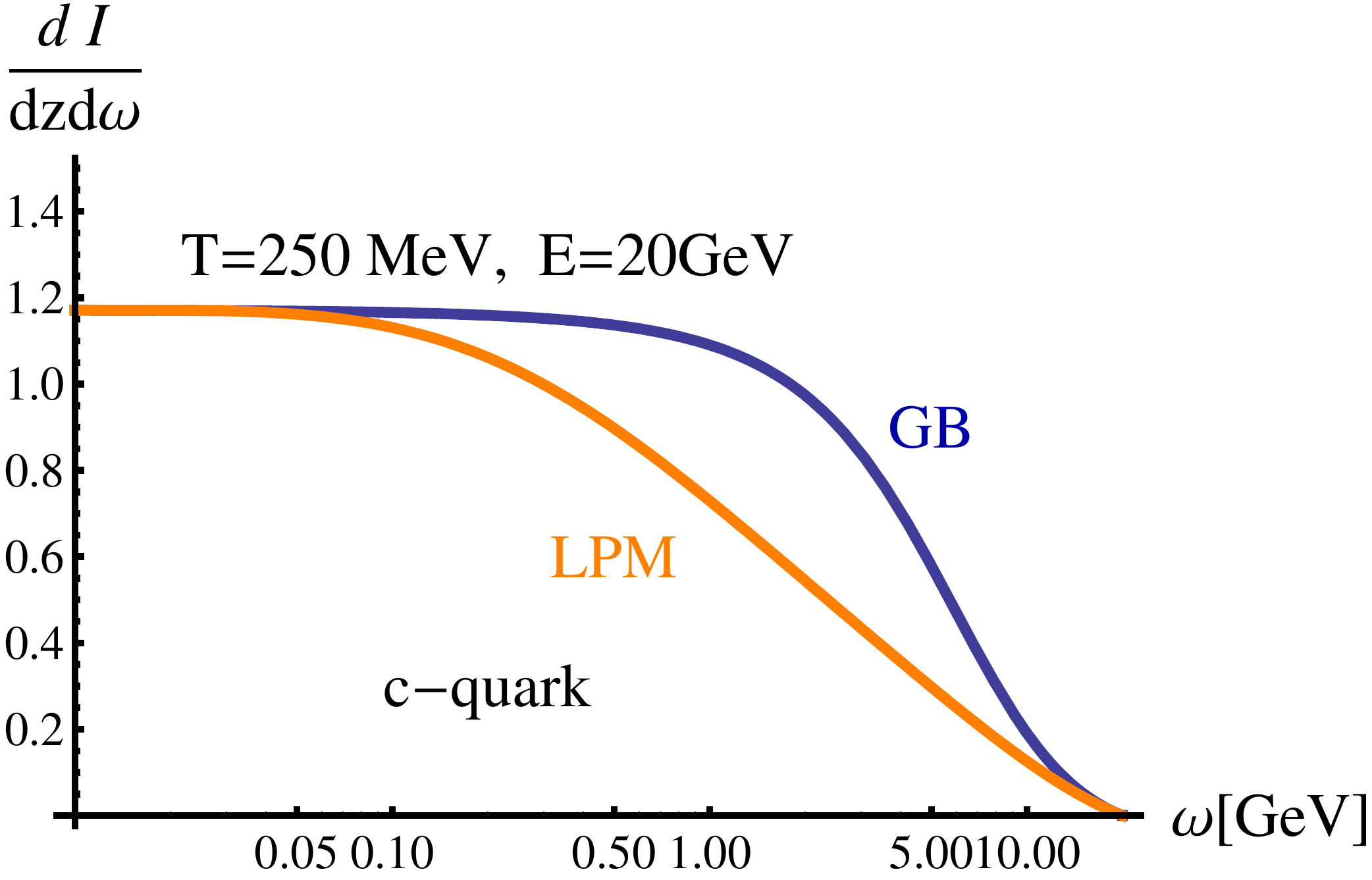}
\hspace{1cm} 
\includegraphics[height=3.cm] {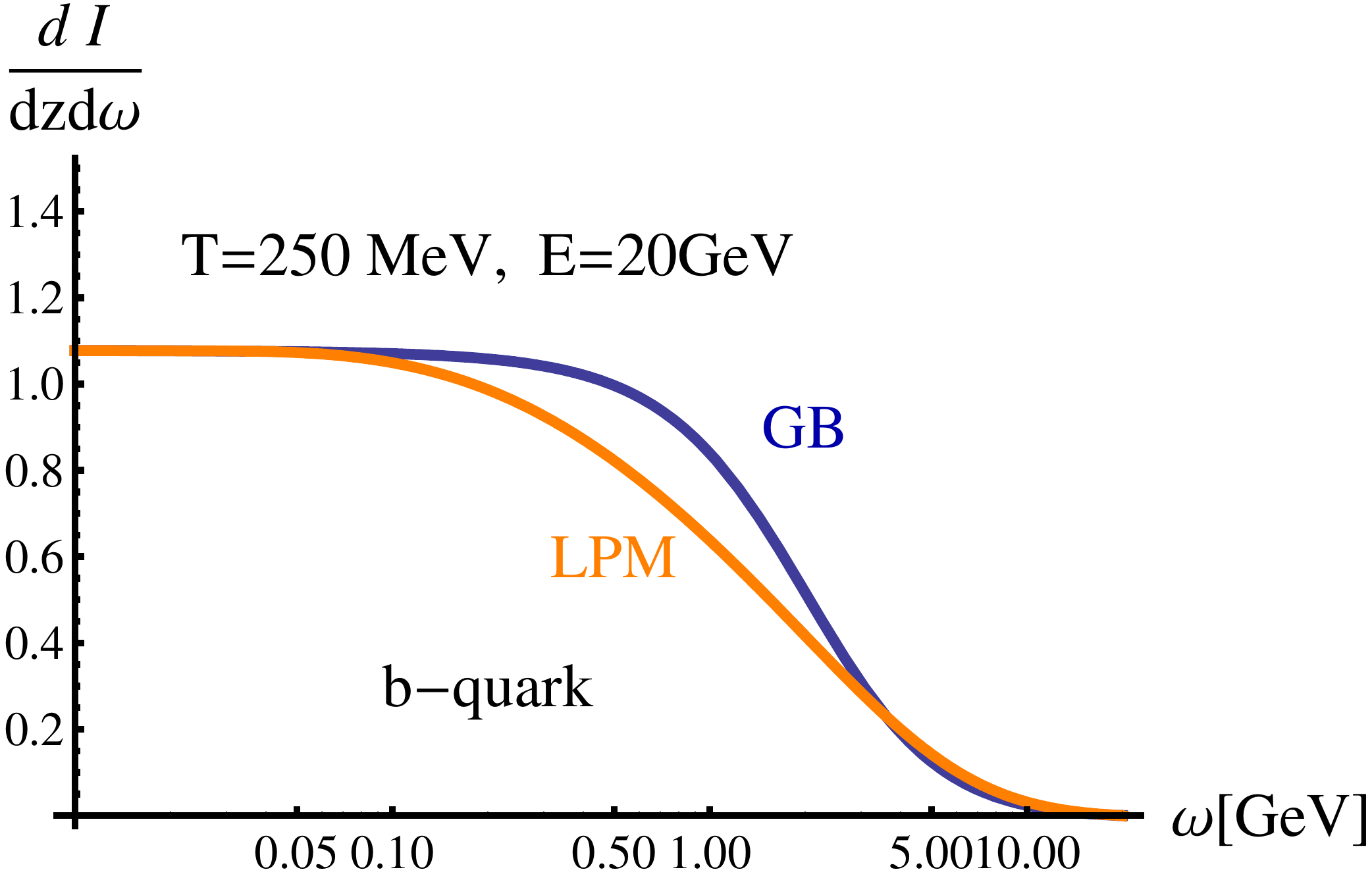}
\end{center}
\caption{Power spectra per unit length for gluon emission through a single scattering (GB) and
for gluon emission through a multiple scattering process modeled by a single effective scatterer
(LPM); see text for details.} 
\label{fig_dI}
\end{figure}

Let us recall that one obtains the power spectrum of gluons induced by single scatterings as the ratio 
$\frac{1}{\sigma_{\rm el}} \; x\frac{d\sigma^{qQ\rightarrow qQg}}{dx}$, that is following relation (12) 
of \cite{Gossiaux:sqm09}: 
\begin{equation}
x\frac{dN_g}{dx}=\frac{2 N_c \alpha_s (1-x)}{\pi} \int d^2q_\perp
\ln\left(1+\frac{q_\perp^2}{3 (m_g^2+x^2 M^2)}\right) {\cal P}_1(q_\perp)\,,
\quad{\rm where}\quad {\cal P}_1(q_\perp)=\frac{\frac{{d\sigma}^{qQ\rightarrow qQ}_{\rm el}}{d^2q_\perp}}
{\sigma^{qQ\rightarrow qQ}_{\rm el}}
\label{gluon_spectra}
\end{equation}
is the probability density associated with a single (Rutherford) scattering. 
If we now consider the gluon radiated after some rescattering by the medium, as illustrated in fig.
\ref{fig_GB} (right), our {\em basic hypothesis} 
is that the $quark-gluon$ system cannot resolve the various kicks $q_1$, $q_2$, 
\ldots $q_n$ during the formation length, so that all scattering centers act as an effective single one 
shocking the system with a transverse momentum $\vec{Q}_\perp=\vec{q}_{\perp 1}+\cdot +
\vec{q}_{\perp n}$. Hence, we could extend  eq. (\ref{gluon_spectra}) by substituting 
${\cal P}_1(q_\perp)\rightarrow {\cal P}_{n=\bar{N}_{\rm coh}}(Q_\perp)$, where 
$\bar{N}_{\rm coh}=\frac{l_f}{\tilde{\lambda}}$ is the average ``coherence number'' and ${\cal P}_{n}$
is obtained from ${\cal P}_{1}$ through convolution. 
However, it is important to note that the multi-scattering process causes the radiation to happen in 
a shorter formation length and cures the collinear divergency that would appear for massless partons. Those
two facts are correlated as the amplitude of the gluon field is $\propto l_f$~\cite{Sorensen:92}, so that 
the differential power spectrum $dI\propto l_f^2$. More precisely, the finite-mass GB transition 
probability writes
\begin{equation}
|M_{\rm GB}|^2\propto \left|\frac{\vec{k}_\perp}{k_\perp^2 +x^2 M^2}-
\frac{\vec{k}_\perp-\vec{q}_\perp}{(\vec{k}_\perp-\vec{q}_\perp)^2 +x^2 M^2}\right|^2
\propto \left|l_{f,{\rm sing}}^{3\;{\rm pre}}\vec{k}_\perp-
l_{f,{\rm sing}}^{3\;{\rm gl}}\left(\vec{k}_\perp-\vec{q}_\perp\right)\right|^2\,,
\end{equation}
so that the denominator in the ln of eq. (\ref{gluon_spectra}) can be seen as a reminiscent 
of the collinear formation length in our finite gluon mass approach: 
$m_g^2 + x^2 M^2=\frac{2\omega}{l_{f,{\rm sing}}}$. Based on this interpretation, 
we propose to replace the denominator in the ln by a quantity evolving smoothly from 
$\frac{2\omega}{l_{f,{\rm sing}}}\rightarrow \frac{2\omega}{l_{f,{\rm mult}}}$ when coherence effects set in: 
\begin{equation}
m_g^2 + x^2 M^2=\frac{2\omega}{l_{f,{\rm sing}}} \rightarrow 
\tilde{m}_g^2\triangleq 2\omega \left(\frac{1}{l_{f,{\rm sing}}}+
\frac{1}{l_{f,{\rm mult}}}\right)= m_g^2 + x^2 M^2 + \frac{\bar{N}_{\rm coh}}{2\Phi_{\rm dec}}\langle 
q_\perp^2\rangle=m_g^2 + x^2 M^2 + \sqrt{\frac{\hat{q}\omega}{\Phi_{\rm dec}}}\,.
\end{equation}
Performing these changes in eq. (\ref{gluon_spectra}) and dividing by the formation length $l_f=\bar{N}_{\rm coh}
\tilde{\lambda}$,
one obtains an effective power spectrum per unit length at small $x$, valid for small or large 
$\bar{N}_{\rm coh}$:
\begin{equation}
\frac{d^2I^{x\ll 1}_{\rm model}}{dz\,d\omega}
\approx \frac{2 N_c \alpha_s}{\pi l_f}
\left\langle \ln\left(1+\frac{Q_\perp^2}{3\tilde{m}_g^2}\right)
\right\rangle_{{\cal P}_{\bar{N}_{\rm coh}}}\,.
\label{spectre_eff_rayonnement_per_length_mult_base_qcd}
\end{equation}
Starting from this expression, one can recover all regimes identified in section \ref{sect_estimates_lf}.
In particular, for $\bar{N}_{\rm coh}>1$ but not too large and $\langle q_\perp^2\rangle \ll \tilde{m}_g^2$, one
finds $\frac{1}{\bar{N}_{\rm coh}}\left\langle \ln\left(1+\frac{Q_\perp^2}{3\tilde{m}_g^2}\right) 
\right\rangle = {\cal O}(\bar{N}_{\rm coh}^{0})$ which indicates that coherence effects do not alter the spectrum 
significantly. This corresponds to the case $\lambda<l_f< L^{\star\star}$ discussed above. To calibrate
our model, we consider the analytical results established in \cite{Baier:97} for $m_g=M=0$ and 
$\bar{N}_{\rm coh}\gg 1$:
\begin{equation}
\frac{d^2I_{\rm BDMPS}}{dz d\omega}=\frac{3C_F}{\pi} \times \frac{\alpha_s}{\tilde{\lambda}}
\sqrt{\tilde{\kappa} \ln\frac{1}{\tilde{\kappa}}}\quad
{\rm where}\quad \tilde{\kappa}=\frac{\tilde{\lambda}\mu^2}{2\omega}\,.
\label{eq_bdmps_d2idzdom}
\end{equation}
To make the connexion to this result, one has to consider more seriously the case of Coulomb scatterings, 
for which ${\cal P}_1\propto 
\frac{1}{(q_\perp^2+\mu^2)^2}$. For a given $\bar{N}_{\rm coh}\gg 1$, one finds~\cite{Peigne:09} that
$\langle Q_\perp^2\rangle \propto \bar{N}_{\rm coh} \sqrt{\bar{N}_{\rm coh}}\,\mu^2$ instead of
$\langle Q_\perp^2\rangle \propto \bar{N}_{\rm coh} \mu^2$ for Gaussian diffusion. This leads to an 
effective $\hat{q}=\frac{\mu^2}{\tilde{\lambda}}\times \ln\left(2\sqrt{\frac{\omega \Phi_{\rm dec}}
{\tilde{\lambda}\mu^2}}\right)$ and a reduction of the formation length 
$l_{f}\rightarrow l_f^{\rm coul}=l_f/\sqrt{\ln\left(2\sqrt{\frac{\omega \Phi_{\rm dec}}
{\tilde{\lambda}\mu^2}}\right)}$ in our expression (\ref{spectre_eff_rayonnement_per_length_mult_base_qcd})
which then shows the $\sqrt{\tilde{\kappa} \ln\frac{1}{\tilde{\kappa}}}$ dependence of eq. 
(\ref{eq_bdmps_d2idzdom}), with
\begin{equation}
\frac{d^2I_{\rm model}}{d^2I_{\rm BDMPS}}=\frac{N_c}{3 C_F}\,
\frac{\ln(1+\frac{2\Phi_{\rm dec}}{3})}{\sqrt{\Phi_{\rm dec}}}\,.
\end{equation}
The first factor reflects the difference between the color factor associated to the planar diagrams 
that dominates in the large $N_c$ large $\bar{N}_{\rm coh}$ limit~\cite{Baier:97} and the color 
factors in a single process while the second reflects the ``quality'' of our effective scattering
center approach. For practical purposes, we choose $\Phi_{\rm dec}=2$ and rescale our model     
(\ref{spectre_eff_rayonnement_per_length_mult_base_qcd}) by a factor $\frac{3 C_F}{0.6 N_c}=\frac{5 C_F}{N_c}$ 
when $\bar{N}_{\rm coh}\gg 1$, in order to guarantee quantitative agreement with
(\ref{eq_bdmps_d2idzdom}); we also evaluate ${\cal P}_1$ as well as the transport coefficient $\hat{q}$ entering
our expressions with the one effective gluon exchange model presented in \cite{Gossiaux:08}. 
In fig.~\ref{fig_dI}, we display some radiation spectra per unit length resulting from our model for typical 
parameters encountered in URHIC.
For $b$-quarks, $20\;{\rm GeV}/c\approx E_{\rm no\; LPM}$. Thus, one is still 
in the low energy regime and impact of coherence on the spectrum is pretty weak; 
for $c$-quarks, $10~{\rm GeV}/c$ and $20~{\rm GeV}/c$ belong to the intermediate energy regime. 
The effect is more pronounced and increases with $E$ as expected. 

\section{Consequences for heavy quark quenching in 
ultrarelativistic heavy ion collisions and conclusion}
\begin{figure}[H]
\begin{center}
\includegraphics[height=4cm] {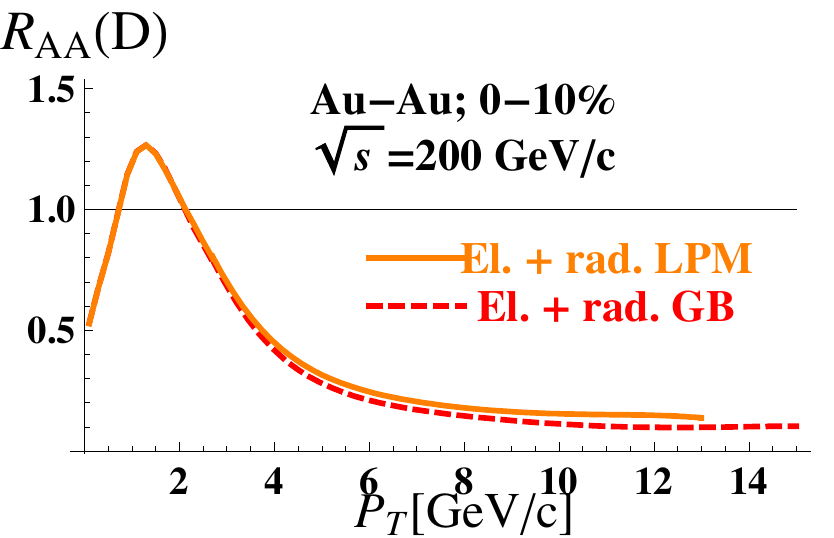}
\hspace{1cm} 
\includegraphics[height=4cm] {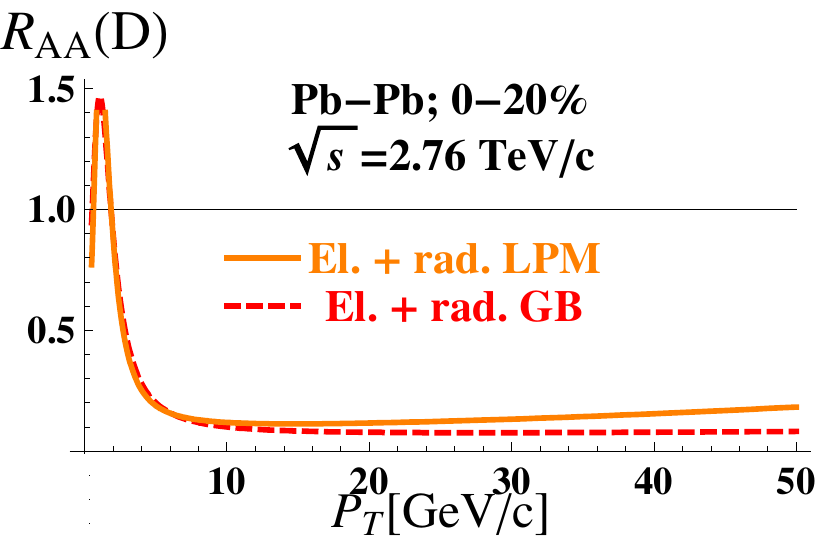} 
\end{center}
\caption{Dashed: $R_{AA}$ of $D$-mesons obtained with our MC$\alpha_s$HQ generator 
\cite{Gossiaux:08} with elastic + radiative energy loss evaluated by extending the GB approach to massive quarks; 
plain: same with coherence effect included for radiative energy loss.}
\label{fig_RAAD}
\end{figure}

In fig. \ref{fig_RAAD}, we present the nuclear modification factor ($R_{AA}$) of $D$-mesons 
in central Au-Au (RHIC) and Pb-Pb (LHC) collisions. The ``El. + rad. GB'' curves correspond to the 
calculation presented in~\cite{Gossiaux:sqm09}, with $\alpha_s=0.3$ in eq. ($\ref{gluon_spectra}$) and a 
rescaling of the interaction rate by a factor $K=0.7$ introduced to reproduce the $R_{AA}$ of non
photonic single electrons measured at RHIC. The ``El. + rad. LPM'' curves correspond to 
the model presented in section~\ref{section_model}. One observes a typical increase of 40\% (resp. 100\%) 
due to coherence for the $R_{AA}(D)$ at $p_T=10~{\rm GeV/c}$ (resp. $p_T=50~{\rm GeV/c}$). This 
originates from the reduced quenching in the ``rad. LPM'' case (see fig. \ref{fig_dI}), which is also responsible 
for the increase of $R_{AA}$ with $p_T$ for $p_T\gtrsim 15~{\rm GeV}/c$. We conclude that the $p_T$-range 
available at LHC offers better chance to discriminate between models of HQ energy loss, although it will require 
good experimental accuracy.





\bibliographystyle{elsarticle-num}
\bibliography{<your-bib-database>}



\end{document}